\documentclass[journal]{IEEEtran}

\usepackage[utf8]{inputenc}
\usepackage[T1]{fontenc}
\usepackage{graphicx}
\usepackage{array}
\usepackage{multirow}
\usepackage{url}
\usepackage{cite}
\usepackage{amsmath}
\usepackage{xcolor}
\usepackage{tikz}
\usetikzlibrary{arrows.meta, fit, positioning}

\colorlet{nnconv}{black!6}
\colorlet{nndense}{black!16}
\colorlet{nnout}{black!30}
\tikzset{
  nnarr/.style={-{Stealth[length=1.4mm]}, semithick},
  nnnote/.style={font=\tiny, align=center, inner sep=1pt}}
\newcommand{\nnlayer}[5]{%
  \node[draw, minimum width=#3, minimum height=2.4cm, inner sep=0pt, #4, #2]
    (#1) {};
  \node[rotate=90, align=center, font=\scriptsize, inner sep=0pt] at (#1)
    {#5};}
\usepackage[hidelinks]{hyperref}

\begin{document}

\title{A Multimodal Emotion Recognition System:\\ Integrating Facial
  Expressions, Body Movement, Speech, and Spoken Language}

\author{Kris~Kraack%
\thanks{K. Kraack is with the College of Computing, Georgia Institute of
  Technology, Atlanta, GA, United States of America (e-mail:
  kkraack3@gatech.edu).}}

\maketitle

\begin{abstract}
Psychological evaluation relies on clinicians' observation and
interpretation of a patient's emotional expression, which is susceptible to
subjectivity, bias, fatigue, and inconsistency. This paper presents a
multimodal emotion recognition system designed to support, rather
than replace, the evaluator. The system combines facial expression
recognition, speech emotion recognition, text-based emotion recognition of
transcribed speech, and body movement analysis, and fuses their outputs into
an emotional profile of class probabilities that is presented to the
evaluator together with the overall prediction. Although the unimodal models
achieve high accuracy on their benchmark test sets, a remote user study with
52 participants expressing seven emotional states shows that their accuracy
falls to between 82.97\% and 87.64\% under realistic conditions, with
individual emotional states recognised in as few as 48.08\% of cases. Under
the same conditions, the multimodal system achieves an accuracy of 96.43\%,
significantly higher than every single-modality component
($p < 10^{-4}$), with no emotional state below 92.31\%. These results indicate that strong
benchmark performance of a single channel does not transfer to realistic use,
and that multimodal fusion provides the robustness required of a tool that
supports psychological evaluation.
\end{abstract}


% =====================================================================
\section{Introduction}
\label{sec:introduction}

\IEEEPARstart{P}{sychological} evaluation depends on the clinician's reading
of a patient's emotional state from what is said, how it is said, and how the
face and body move. This judgement is made by human evaluators and is
therefore subject to subjective interpretation and bias, which can lead to
inconsistent and inaccurate diagnoses~\cite{b1}. Evaluators are further
affected by attention and compassion fatigue, perceptual biases, and
imperfect recall~\cite{b2,b3}. A quantitative, data-driven aid that offers a
second, independent reading of emotional state could help to reduce these
sources of error.

This paper presents a multimodal emotion recognition system intended as such
an aid for psychologists, psychiatrists, and other trained evaluators. The
system analyses four channels: facial expression, body movement, the acoustic
properties of speech, and the linguistic content of what is said. Facial
expression recognition (FER) uses a convolutional neural network (CNN), body
movement is analysed through pose estimation, speech emotion recognition
(SER) uses a CNN on acoustic features, and text-based emotion recognition
applies a bidirectional long short-term memory (BiLSTM) network to
automatically transcribed speech. A fusion stage combines the outputs of the
components into an emotional profile, which is presented to the evaluator
together with the overall prediction and the activity level. The system is intended to corroborate
or challenge the evaluator's impression, not to replace clinical judgement.

The central finding of this work concerns the gap between benchmark and
realistic performance. The objective is not to achieve state-of-the-art
results, but to show that strong benchmark performance does not guarantee
reliability in practice. None of the single-modality components exceeds the
state of the art, but each performs well on its benchmark test set. Under
realistic conditions, however, a single channel can be absent, ambiguous, or
misleading, and its accuracy falls accordingly, particularly for specific
emotional states. Combining channels allows the system to compensate
for the weaknesses of each, and the multimodal system outperforms every
single-modality component under identical conditions.

The contributions of this paper are as follows:
\begin{itemize}
  \item A multimodal emotion recognition system for supporting
    psychological evaluation, whose fusion stage presents both the emotional
    profile and the overall prediction to the evaluator.
  \item A remote user study in which each single-modality component and the
    multimodal system are evaluated by the same participants on the same
    emotional states.
  \item Evidence that single-modality performance on benchmark test sets does
    not transfer to realistic conditions, and that multimodal fusion
    substantially improves both overall accuracy and consistency across
    emotional states.
\end{itemize}

The remainder of this paper is organised as follows.
Section~\ref{sec:background} outlines the limitations of human psychological
evaluation that motivate this work, and Section~\ref{sec:related} reviews
related work. Section~\ref{sec:system} describes the system and its
components. Section~\ref{sec:evaluation} reports the performance of the
components on benchmark datasets, and Section~\ref{sec:study} presents the
user study. Section~\ref{sec:discussion} discusses the findings, limitations,
and future work, and Section~\ref{sec:conclusion} concludes the paper.

% =====================================================================
\section{Limitations of Human Psychological Evaluation}
\label{sec:background}

Clinical judgement is susceptible to biases arising from cultural, gender, and
socioeconomic factors, which affect the objectivity of
assessment~\cite{b2,b3}. Evaluators may also find it difficult to recognise
when patients, consciously or unconsciously, conceal or misrepresent their
emotional state. Much of the relevant information is carried by subtle and
fleeting non-verbal cues, such as shifts in posture, changes in speech
patterns, eye contact, and micro-expressions lasting only fractions of a
second, which are difficult for human observers to detect, particularly under
time pressure or in stressful conditions~\cite{b27}. Misreading such cues can
lead to erroneous conclusions.

Psychological tests are typically administered within a short, scheduled
window, so that the observed presentation may reflect transient circumstances,
such as recent adverse events or acute stressors, rather than the patient's
typical state~\cite{b28}. Continuous recognition of emotional state across
an interview could help to identify presentations that are unusual for a
particular patient before formal testing is conducted.

Attention and compassion fatigue, induced by stress and workload, further
reduce diagnostic accuracy and increase the frequency of
errors~\cite{b29,b30,b31}. Evaluations also depend on the clinician's recall
of the session, which is filtered by memory and attention, and rising
diagnosis rates add further pressure to healthcare systems~\cite{b32}. The
consequences of these limitations are borne by patients: misdiagnosis can
lead to inappropriate treatment and delayed access to appropriate
care~\cite{b33}, while overdiagnosis contributes to unnecessary treatment and
stigma~\cite{b34}.

Technology has been proposed as a means of mitigating these
limitations~\cite{b35,b36}. A system that analyses several modalities can
provide a quantitative complement to the evaluator's judgement, drawing
attention to cues that may have been missed and reinforcing or questioning
initial observations. For its observations to be consistent, the recording
conditions must be standardised, including the recording equipment, the
positioning of the participant, background interference, camera placement and
angle, lighting, and microphone placement. Such a system cannot replace the
understanding and empathy of a human evaluator; its role is to support the
evaluator's decision-making.

% =====================================================================
\section{Related Work}
\label{sec:related}

Facial expression recognition has been studied
extensively~\cite{b4,b5,b6,b7}, with deep learning models, in particular
CNNs, now the dominant approach. Pise et al.~\cite{b8} emphasise the role of
facial expressions within multimodal emotion recognition. Speech emotion
recognition~\cite{b9,b10,b11,b12} and emotion recognition from
text~\cite{b13,b14,b15} are similarly well established and contribute
complementary acoustic and linguistic cues. Multimodal emotion recognition
systems that combine these or related modalities have been proposed in
several studies~\cite{b16,b17,b18,b19,b20,b21}, which report that combining
modalities can overcome limitations of unimodal approaches. Wagner et
al.~\cite{b21} further examine fusion methods that remain robust when
modalities are missing.

The present work differs from these systems in three respects. First, the
system and its recording environment are designed specifically for
psychological evaluation: participants are recorded against a static
background, and the training images were cropped to the face, which removes
background interference. Second, the fusion stage presents the emotional
profile, that is, the predicted probability of each emotional state, together
with the overall prediction and the activity level from the body movement
component, rather than a single label. Third, the system
is evaluated in a user study that compares each single-modality component with
the multimodal system on the same participants and emotional states.

% =====================================================================
\section{System Description}
\label{sec:system}

The system comprises four components, for facial expression recognition,
body movement analysis, speech emotion recognition, and text-based emotion
recognition, and a fusion stage that combines their outputs
(Fig.~\ref{fig:architecture}). The facial expression and body movement
components analyse the camera feed in real time, while the speech and
text-based components analyse an audio recording of the session.

\begin{figure}[!t]
  \centering
  \begin{tikzpicture}[
      font=\footnotesize,
      box/.style={draw, rectangle, align=center, inner sep=2pt},
      comp/.style={box, minimum width=3.0cm, minimum height=0.95cm},
      arr/.style={-{Stealth[length=1.6mm]}, semithick}]
    % Single-modality components
    \node[comp] (fer)  at (0, 1.65)
      {\textbf{Facial expression}\\{\scriptsize face detection, CNN}};
    \node[comp] (body) at (0, 0.55)
      {\textbf{Body movement}\\{\scriptsize pose estimation}};
    \node[comp] (ser)  at (0,-0.55)
      {\textbf{Speech emotion}\\{\scriptsize acoustic features, CNN}};
    \node[comp] (slr)  at (0,-1.65)
      {\textbf{Text-based emotion}\\{\scriptsize speech-to-text, BiLSTM}};
    % Inputs
    \node[box, minimum width=1.55cm, minimum height=2.05cm] (cam) at (-2.7, 1.1)
      {Camera\\video};
    \node[box, minimum width=1.55cm, minimum height=2.05cm] (mic) at (-2.7,-1.1)
      {Microphone\\audio};
    % Fusion and outputs
    \node[box, minimum width=0.9cm, minimum height=4.25cm] (fus) at (2.4, 0) {};
    \node[rotate=90] at (fus) {Multimodal fusion};
    \node[box, text width=1.6cm, minimum height=0.95cm] (prof) at (4.15, 1.3)
      {Emotional profile\\{\scriptsize class probabilities}};
    \node[box, text width=1.6cm, minimum height=0.8cm] (pred) at (4.15, 0.1)
      {Overall\\prediction};
    \node[box, text width=1.6cm, minimum height=0.8cm] (act) at (4.15,-1.0)
      {Activity\\level};
    \node[font=\scriptsize, text=black!70] (lbl) at (4.15,-1.75)
      {shown to evaluator};
    \node[draw, dashed, black!60, inner sep=3pt, fit=(prof)(pred)(act)(lbl)] {};
    % Connections
    \draw[arr] (cam.east |- fer.west)  -- (fer.west);
    \draw[arr] (cam.east |- body.west) -- (body.west);
    \draw[arr] (mic.east |- ser.west)  -- (ser.west);
    \draw[arr] (mic.east |- slr.west)  -- (slr.west);
    \foreach \c in {fer, body, ser, slr}
      \draw[arr] (\c.east) -- (\c.east -| fus.west);
    \draw[arr] (fus.east |- prof.west) -- (prof.west);
    \draw[arr] (prof.south) -- (pred.north);
    \draw[arr] (fus.east |- act.west) -- (act.west);
  \end{tikzpicture}
  \caption{System architecture. Each component processes its input channel
    independently. The multimodal fusion model combines the outputs of the
    emotion recognition components into an emotional profile of class
    probabilities, from which the overall prediction is derived, and presents
    them together with the activity level from the body movement component.
    All fusion outputs are shown to the evaluator.}
  \label{fig:architecture}
\end{figure}

\subsection{Facial Expression Recognition}

The facial expression recognition component was trained on two datasets:
FER-2013~\cite{b37} and the Real-world Affective Faces Database
(RAF-DB)~\cite{b38}. A model was also trained on the Extended Cohn-Kanade
(CK+) dataset~\cite{b39} and achieved near-perfect performance; however, the
small size and limited diversity of CK+, with only 184 test samples, make it
unlikely to generalise to realistic conditions, and it was therefore not included
in the system. The training data were filtered to match the intended
recording environment, removing samples that do not meet its requirements
through face detection, cropping, and blur filtering
(Table~\ref{tab:fer-data}).

For FER-2013, faces were detected with the Haar cascade classifier~\cite{b40},
three cropping strategies (tight, moderate, and original) were applied to
increase sample diversity, and low-quality images were removed with a Laplacian
blur filter. For RAF-DB, faces were detected with the Dlib face
detector~\cite{b41} and its 68-point facial landmark predictor~\cite{b42},
combined with the same blur filter. Both datasets were augmented with variations
in brightness, rotation, and horizontal flipping using the TensorFlow
ImageDataGenerator; for the model trained on both datasets, a custom function
combined the augmentations of the two datasets per emotion label.

The CNN (Fig.~\ref{fig:fer-cnn}) comprises five convolutional layers
arranged in four blocks, starting with layers of 32 and 64 filters. Each
block uses batch normalisation, max-pooling, and dropout, and the last two
blocks apply L2 weight regularisation. The convolutional blocks are followed
by a dense layer of 128 units with ReLU activation, batch normalisation, and
dropout, and a softmax output layer over seven emotional states. The model
was trained with the Adam optimiser and categorical cross-entropy loss,
separately on FER-2013 and RAF-DB and on their combination.

\begin{figure}[!t]
  \centering
  \begin{tikzpicture}
    \nnlayer{in}{at={(0,0)}}{0.8cm}{}{Face image\\$48{\times}48{\times}1$}
    \nnlayer{c1}{right=0.4cm of in}{0.8cm}{fill=nnconv}
      {Conv 32, $3{\times}3$\\Conv 64, $3{\times}3$}
    \nnlayer{c2}{right=0.4cm of c1}{0.8cm}{fill=nnconv}
      {Conv 128, $5{\times}5$}
    \nnlayer{c3}{right=0.4cm of c2}{0.8cm}{fill=nnconv}
      {Conv 256, $3{\times}3$\\L2}
    \nnlayer{c4}{right=0.4cm of c3}{0.8cm}{fill=nnconv}
      {Conv 512, $3{\times}3$\\L2}
    \nnlayer{fc}{right=0.4cm of c4}{0.8cm}{fill=nndense}
      {Flatten\\Dense 128, BN}
    \nnlayer{out}{right=0.4cm of fc}{0.8cm}{fill=nnout}
      {Softmax\\7 states}
    \foreach \src/\dst in {in/c1, c1/c2, c2/c3, c3/c4, c4/fc, fc/out}
      \draw[nnarr] (\src) -- (\dst);
    % Pooling size and dropout rate per block
    \node[nnnote, below=2pt of in] {pool\\drop};
    \foreach \lyr/\rate in {c1/0.45, c2/0.45, c3/0.55, c4/0.55}
      \node[nnnote, below=2pt of \lyr] {$2{\times}2$\\\rate};
    \node[nnnote, below=2pt of fc] {--\\0.65};
  \end{tikzpicture}
  \caption{Facial expression recognition CNN. Each convolutional block applies
    ReLU convolution, batch normalisation (BN), max-pooling, and dropout,
    with the pooling size and dropout rate given below the block. L2: L2
    weight regularisation ($\lambda = 0.01$).}
  \label{fig:fer-cnn}
\end{figure}

For real-time recognition, faces in the camera feed are detected with the
Dlib frontal face detector~\cite{b41}, and predictions are stabilised with a
frame-based queue. The component outputs a probability distribution over the
emotional states.

\begin{table}[!t]
  \renewcommand{\arraystretch}{1.2}
  \caption{Facial Expression Datasets Before and After Filtering}
  \label{tab:fer-data}
  \centering
  \footnotesize
  \begin{tabular*}{\columnwidth}{@{\extracolsep{\fill}}lcc}
    \hline
    Dataset & Samples, raw (filtered) & Classes, raw (used) \\
    \hline
    FER-2013 & 35,887 (25,041) & 7 (7) \\
    RAF-DB   & 15,339 (11,549) & 7 (7) \\
    CK+      & 920 (920)       & 8 (7) \\
    \hline
  \end{tabular*}
\end{table}

\subsection{Body Movement Analysis}

The body movement component uses the MediaPipe pose estimation
model~\cite{b43} to extract joint positions from the camera feed, track
individual body parts, and compute movement metrics. From these metrics, it
classifies the overall level of physical activity as low, medium, or high,
and reports the body part with the greatest movement. Its output is passed to
the multimodal fusion, where it contributes the behavioural context of the
participant's physical activity.

\subsection{Speech Emotion Recognition}

The speech emotion recognition component was trained on 39,458 audio
recordings compiled from six datasets: the Crowd-Sourced Emotional Multimodal
Actors Dataset (CREMA-D)~\cite{b44}, the Ryerson Audio-Visual Database of
Emotional Speech and Song (RAVDESS)~\cite{b45}, the Surrey Audio-Visual
Expressed Emotion (SAVEE) database~\cite{b46}, the Toronto Emotional Speech
Set (TESS)~\cite{b47}, the Emotional Speech Dataset (ESD)~\cite{b48,b49}, and
the Multimodal EmotionLines Dataset (MELD)~\cite{b50,b51}, from whose video
files the audio was extracted. The emotion labels of the datasets were
harmonised into seven emotional states.

The recordings were divided into training (72\%), validation (8\%), and test
(20\%) sets, stratified by emotional state, before any augmentation, so that
no recording or augmented version of it appears in more than one set. The
zero-crossing rate, root mean square energy, and Mel-frequency cepstral
coefficients were extracted for each frame of each recording, so that the
input to the model is a sequence of frames with these features as channels. To increase robustness to
recording conditions, five augmented versions of each training recording were
added: with added noise, with dynamic range compression, with both, with pitch
shifting, and with pitch shifting and noise. The validation and test sets
contain only the original recordings.

The CNN (Fig.~\ref{fig:ser-cnn}) comprises seven convolutional layers,
beginning with a block of 512
filters with a kernel size of 5, followed by progressively deeper
convolutional blocks, each with batch normalisation and pooling, and dropout
at several stages. A dense layer of 512 units with batch normalisation
precedes the softmax output layer over seven emotional states. The model was
trained with the Adam optimiser, categorical cross-entropy loss, and balanced
class weights, with early stopping and learning-rate reduction on the
validation set. At inference, the component processes the audio recording of the session and
outputs a probability distribution over the emotional states.

\begin{figure}[!t]
  \centering
  \begin{tikzpicture}
    \nnlayer{in}{at={(0,0)}}{0.8cm}{}{ZCR, RMS, MFCC\\frames${}\times{}$features}
    \nnlayer{c1}{right=0.28cm of in}{0.5cm}{fill=nnconv}{Conv1D 512, $k{=}5$}
    \nnlayer{c2}{right=0.28cm of c1}{0.5cm}{fill=nnconv}{Conv1D 512, $k{=}5$}
    \nnlayer{c3}{right=0.28cm of c2}{0.5cm}{fill=nnconv}{Conv1D 512, $k{=}3$}
    \nnlayer{c4}{right=0.28cm of c3}{0.5cm}{fill=nnconv}{Conv1D 256, $k{=}3$}
    \nnlayer{c5}{right=0.28cm of c4}{0.5cm}{fill=nnconv}{Conv1D 256, $k{=}3$}
    \nnlayer{c6}{right=0.28cm of c5}{0.5cm}{fill=nnconv}{Conv1D 128, $k{=}3$}
    \nnlayer{c7}{right=0.28cm of c6}{0.5cm}{fill=nnconv}{Conv1D 128, $k{=}3$}
    \nnlayer{fc}{right=0.28cm of c7}{0.8cm}{fill=nndense}
      {Flatten\\Dense 512, BN}
    \nnlayer{out}{right=0.28cm of fc}{0.8cm}{fill=nnout}{Softmax\\7 states}
    \foreach \src/\dst in {in/c1, c1/c2, c2/c3, c3/c4, c4/c5, c5/c6, c6/c7,
                           c7/fc, fc/out}
      \draw[nnarr] (\src) -- (\dst);
    % Pooling size and dropout rate per block
    \node[nnnote, below=2pt of in] {pool\\drop};
    \foreach \lyr/\pool/\rate in {c1/2/--, c2/2/0.2, c3/3/--, c4/3/0.15,
                                  c5/2/0.15, c6/2/0.15, c7/2/0.15}
      \node[nnnote, below=2pt of \lyr] {\pool\\\rate};
    \node[nnnote, below=2pt of fc] {--\\--};
  \end{tikzpicture}
  \caption{Speech emotion recognition CNN. The input is the sequence of
    frames of a recording, each described by its zero-crossing rate (ZCR),
    root mean square energy (RMS), and Mel-frequency cepstral coefficients
    (MFCC); the convolutions run along the time axis. Each
    convolutional block applies ReLU convolution with kernel size $k$, batch
    normalisation (BN), max-pooling with stride 2, and dropout, with the
    pooling size and dropout rate given below the block.}
  \label{fig:ser-cnn}
\end{figure}

\subsection{Text-Based Emotion Recognition}

The text-based component recognises emotion from the linguistic content of
speech. It was trained on four text datasets: GoEmotions~\cite{b52}, the
Emotion Classification Dataset (ECD)~\cite{b53}, the Emotions Dataset
(ED)~\cite{b54}, and Sentiment140~\cite{b55}. The texts were normalised with
regular expressions, and a keyword dictionary was constructed from the most
frequent terms for each of six emotional states (anger, disgust, joy, sadness,
surprise, and fear), after removing standard and custom stop words. The texts
were then labelled with this dictionary (Table~\ref{tab:nlp-data}).

Duplicate texts were removed, leaving 806,530 unique texts, which were
divided into training (72\%), validation (8\%), and test (20\%) sets. The
texts were tokenised and padded to a fixed length. The model
(Fig.~\ref{fig:nlp-model}) consists of an embedding layer, a BiLSTM layer with 256 units, batch
normalisation and dropout, a dense layer of 128 units, and a softmax output
layer over the six emotional states. When no state reaches a predicted
probability of 50\%, no specific emotion is assigned and the result is
treated as neutral. The model was trained with the Adam optimiser and sparse
categorical cross-entropy loss, with early stopping and learning-rate
reduction on the validation set. At inference, the audio recording is
transcribed with the Whisper speech-to-text model~\cite{b56}, and the
transcript is classified to give a probability distribution over the
emotional states.

\begin{figure}[!t]
  \centering
  \begin{tikzpicture}
    \nnlayer{in}{at={(0,0)}}{0.8cm}{}{Token sequence\\178 tokens}
    \nnlayer{emb}{right=0.4cm of in}{0.8cm}{fill=nnconv}{Embedding\\256-d}
    \nnlayer{rnn}{right=0.4cm of emb}{0.8cm}{fill=nnconv}
      {BiLSTM\\$2{\times}256$ units}
    \nnlayer{reg}{right=0.4cm of rnn}{0.8cm}{fill=nnconv}{BN\\Dropout 0.5}
    \nnlayer{fc}{right=0.4cm of reg}{0.8cm}{fill=nndense}
      {Dense 128\\Dropout 0.5}
    \nnlayer{out}{right=0.4cm of fc}{0.8cm}{fill=nnout}{Softmax\\6 states}
    \nnlayer{thr}{right=0.4cm of out}{0.8cm}{dashed}
      {Neutral if\\$\max_k p(k){<}0.5$}
    \foreach \src/\dst in {in/emb, emb/rnn, rnn/reg, reg/fc, fc/out, out/thr}
      \draw[nnarr] (\src) -- (\dst);
  \end{tikzpicture}
  \caption{Text-based emotion recognition model. The BiLSTM reads the
    embedded token sequence in both directions with 256 units each; the
    dense layer uses ReLU activation. When no emotional state reaches a
    predicted probability of 0.5, the text is treated as neutral (dashed).}
  \label{fig:nlp-model}
\end{figure}

\begin{table}[!t]
  \renewcommand{\arraystretch}{1.2}
  \caption{Text Datasets Before and After Labelling}
  \label{tab:nlp-data}
  \centering
  \footnotesize
  \begin{tabular*}{\columnwidth}{@{\extracolsep{\fill}}lcc}
    \hline
    Dataset & Samples, raw (labelled) & Classes, raw (target) \\
    \hline
    GoEmotions   & 58,009 (9,716)      & 28 (6) \\
    ECD          & 5,934 (2,688)       & 3 (6) \\
    ED           & 416,809 (382,255)   & 6 (6) \\
    Sentiment140 & 1,600,000 (433,106) & 3 (6) \\
    \hline
  \end{tabular*}
\end{table}

\subsection{Multimodal Fusion}

The multimodal fusion model combines the outputs of all components at the
decision level (Fig.~\ref{fig:architecture}). Let
$\mathbf{p}_m = (p_m(1), \dots, p_m(K))$ denote the class probability vector
produced by emotion recognition component $m \in \mathcal{M}$ (facial
expression, speech, and text) over the $K$ emotional states, and let
$\mathcal{A} \subseteq \mathcal{M}$ be the set of components that are
available for a given observation. The fusion model computes the emotional
profile
\begin{equation}
  \mathbf{p} = \frac{\sum_{m \in \mathcal{A}} w_m \, \mathbf{p}_m}
                    {\sum_{m \in \mathcal{A}} w_m},
  \label{eq:fusion}
\end{equation}
where the weight $w_m$ of each component increases with the confidence of its
prediction, so that a confident channel contributes more to the profile than an
uncertain one. The overall prediction is the emotional state with the highest
fused probability, $\hat{y} = \arg\max_k p(k)$. Because the sums in
\eqref{eq:fusion} run only over the available components, the model handles
missing modalities, for example when no face is detected or no speech is
recorded, without retraining and without imputing the missing outputs.

The emotional profile $\mathbf{p}$ acts as an interpretable intermediate
representation between the components and the final decision. The fusion
output combines the profile, the overall prediction $\hat{y}$, and the
activity level estimated by the body movement component, and all three are
presented to the evaluator,
so that the basis of each prediction, including the extent to which the
channels agree, remains visible rather than being reduced to a single label.

Relying on several channels allows the system to accommodate cases in which
one channel is uninformative or misleading. Individuals with limited facial
expressiveness, for example, may be misjudged by facial expression analysis
alone, and a single modality cannot account for subtle or conflicting cues
present in other channels.

The system was deployed with a simple web-based user interface for the
duration of the user study, allowing participants to use it remotely.

% =====================================================================
\section{Component Evaluation on Benchmark Datasets}
\label{sec:evaluation}

This section reports the performance of the three emotion recognition
components on the held-out test sets of their training data
(Table~\ref{tab:model-summary}). The body movement component, which does not
classify emotional states, and the multimodal fusion are not included: none
of the training datasets contains all modalities for the same sample, so the
fusion cannot be evaluated on them, and it is instead evaluated in the user
study in Section~\ref{sec:study}.

\begin{table}[!t]
  \renewcommand{\arraystretch}{1.2}
  \caption{Benchmark Performance of the Emotion Recognition Components}
  \label{tab:model-summary}
  \centering
  \footnotesize
  \setlength{\tabcolsep}{2pt}
  \begin{tabular*}{\columnwidth}{@{\extracolsep{\fill}}lccccc}
    \hline
    Component & Accuracy & Loss & Precision & Recall & F1 \\
    \hline
    Facial expression (RAF-DB) & 0.8521 & 0.6857 & 0.8504 & 0.8521 & 0.8496 \\
    Speech emotion    & 0.9963 & 0.0124 & 0.9963 & 0.9963 & 0.9963 \\
    Text-based        & 0.9796 & 0.0621 & 0.9801 & 0.9796 & 0.9794 \\
    \hline
  \end{tabular*}
\end{table}

For facial expression recognition, the model trained on FER-2013 achieved an
accuracy of 68.35\% (loss 1.1072, precision 67.92\%, recall 68.35\%, F1
66.80\%), with frequent confusion of sadness, fear, and anger with the neutral
state. The model trained on RAF-DB performed considerably better, with an
accuracy of 85.21\% (loss 0.6857, precision 85.04\%, recall 85.21\%, F1
84.96\%), although fear remained difficult to recognise. The model trained on
both datasets achieved 75.33\% (loss 0.8569, precision 75.58\%, recall
75.33\%, F1 74.64\%), improving on the FER-2013 model but not on the RAF-DB
model, which was therefore used in the system. Per-class results are given in
Table~\ref{tab:fer-metrics} in the appendix.

The speech emotion recognition model achieved an accuracy of 99.63\% (loss
0.0124, precision, recall, and F1 99.63\%). The
text-based model achieved an accuracy of 97.96\% (loss 0.0621, precision
98.01\%, recall 97.96\%, F1 97.94\%).

% =====================================================================
\section{User Study}
\label{sec:study}

A remote user study was conducted to evaluate each component and the
multimodal system under realistic conditions, with the same participants
and emotional states used throughout.

\subsection{Participants}

Of 159 individuals who began the study, 52 completed it. Participants used
their own computer (macOS or Windows) with a webcam and microphone meeting
the minimum specification. No personally identifiable information was
collected; demographic information was obtained through a survey with an
option not to disclose (Table~\ref{tab:demographics}). Informed consent was
obtained from all participants, who were informed that their camera and
microphone would be used for analysis. Camera images were not stored, audio
recordings were stored only for the duration of processing and then deleted
after, and only aggregated results were retained.

\begin{table}[!t]
  \renewcommand{\arraystretch}{1.1}
  \caption{Demographic Composition of the Participants ($n = 52$)}
  \label{tab:demographics}
  \centering
  \footnotesize
  \begin{tabular*}{\columnwidth}{@{\extracolsep{\fill}}llr}
    \hline
    Characteristic & Category & $n$ (\%) \\
    \hline
    \multirow{5}{*}{Age}
      & 18--24 years  & 14 (26.9) \\
      & 25--34 years  & 19 (36.5) \\
      & 35--44 years  & 13 (25.0) \\
      & 45--54 years  & 5 (9.6)   \\
      & 55+ years     & 1 (1.9)   \\
    \hline
    \multirow{4}{*}{Gender}
      & Male          & 25 (48.1) \\
      & Female        & 22 (42.3) \\
      & Other         & 1 (1.9)   \\
      & Undisclosed   & 4 (7.7)   \\
    \hline
    \multirow{5}{*}{Ethnicity}
      & Caucasian     & 22 (42.3) \\
      & Asian         & 9 (17.3)  \\
      & Black         & 5 (9.6)   \\
      & Latino        & 4 (7.7)   \\
      & Undisclosed   & 12 (23.1) \\
    \hline
    \multirow{4}{*}{Region}
      & Europe        & 23 (44.2) \\
      & North America & 19 (36.5) \\
      & Asia-Pacific  & 8 (15.4)  \\
      & Latin America & 2 (3.8)   \\
    \hline
  \end{tabular*}
\end{table}

\subsection{Procedure}

Each participant completed the study in a single session consisting of two
phases. In Phase~1, participants recorded themselves expressing each emotional
state and judged the output of each single-modality component. In Phase~2,
they judged the output of the multimodal system for the same recordings. No
new recordings were made in Phase~2, so the single-modality components and the
multimodal system were judged on identical input. Before the first phase,
participants were guided, through written instructions and example images, to
set up a recording environment that approximates the standardised conditions
of a psychological interview.

\textit{Phase~1 (single-modality components):} Participants recorded
themselves expressing each of seven emotional states (anger, disgust, fear,
happiness, neutral, sadness, and surprise). For each state, they thought of
and talked about a situation corresponding to it, and then recited
pre-defined sentences for the same state, which makes the spoken content
comparable across participants. After each recording, the output of each
single-modality component was shown, and participants indicated whether it
correctly identified the expressed state by answering true or false. For the
body movement component, participants made sure that their whole body was
visible, performed movements of low, medium, and high intensity, and judged
the estimated activity level in the same manner.

\textit{Phase~2 (multimodal system):} The recordings from Phase~1 were
presented again, one emotional state at a time, together with the prediction
of the multimodal fusion model, and participants again answered true or
false; they could replay a recording before answering.

This yields one judgement per participant, emotional state, and system, that
is, 52 judgements per emotional state and 364 per system.

\subsection{Results}

Table~\ref{tab:study-results} reports the accuracy of each emotion
recognition component and of the multimodal system. Among the single-modality
components, text-based emotion recognition achieved the highest overall
accuracy (87.64\%), followed by speech emotion recognition (85.71\%) and
facial expression recognition (82.97\%). These overall figures, however,
conceal substantial weaknesses for particular emotional states. Facial
expression recognition identified fear in only 61.54\% of cases and disgust
in 69.23\%, speech emotion recognition identified surprise in 71.15\%, and
text-based recognition identified the neutral state in only 48.08\%, as
neutral is recognised only through the confidence fallback described in
Section~\ref{sec:system}-D. An evaluator relying on a single channel would
thus misjudge fear in 38.46\% of cases from facial expression alone, surprise
in 28.85\% from the voice alone, and the neutral state in 51.92\% from the
spoken content alone.

\begin{table}[!t]
  \renewcommand{\arraystretch}{1.15}
  \caption{Accuracy (\%) in the User Study per Emotional State. Each Value
    Is Based on 52 Judgements, and Each Overall Value on 364}
  \label{tab:study-results}
  \centering
  \footnotesize
  \begin{tabular*}{\columnwidth}{@{\extracolsep{\fill}}lcccc}
    \hline
    Emotional state & Facial & Speech & Text-based & Multimodal \\
    \hline
    Anger     & 92.31  & 94.23 & 94.23 & \textbf{98.08}  \\
    Disgust   & 69.23  & 80.77 & \textbf{96.15} & \textbf{96.15}  \\
    Fear      & 61.54  & 92.31 & \textbf{96.15} & \textbf{96.15}  \\
    Happiness & \textbf{100.00} & 94.23 & 98.08 & \textbf{100.00} \\
    Neutral   & 84.62  & 82.69 & 48.08 & \textbf{92.31}  \\
    Sadness   & 90.38  & 84.62 & 94.23 & \textbf{98.08}  \\
    Surprise  & 82.69  & 71.15 & 86.54 & \textbf{94.23}  \\
    \hline
    Overall   & 82.97  & 85.71 & 87.64 & \textbf{96.43}  \\
    \hline
  \end{tabular*}
\end{table}

The multimodal system achieved an overall accuracy of 96.43\%, with 13
incorrect judgements out of 364, and recognised every emotional state with an
accuracy of at least 92.31\%. Its weakest state therefore exceeds the overall
accuracy of every single-modality emotion recognition component.

Table~\ref{tab:significance} reports 95\% Wilson confidence intervals for the
overall accuracies and compares the multimodal system with each
single-modality component using Fisher's exact test. The multimodal system is
significantly more accurate than each component ($p < 10^{-4}$, also after
Bonferroni correction for three comparisons), and its confidence interval does
not overlap with that of any component. As only aggregated results were
retained, the tests treat the judgements of the two systems as independent
samples; because the same participants evaluated both, this is a conservative
choice.

\begin{table}[!t]
  \renewcommand{\arraystretch}{1.15}
  \caption{Overall Accuracy in the User Study with 95\% Confidence Intervals
    and Comparison with the Multimodal System}
  \label{tab:significance}
  \centering
  \footnotesize
  \setlength{\tabcolsep}{3pt}
  \begin{tabular*}{\columnwidth}{@{\extracolsep{\fill}}lccc}
    \hline
    System & Accuracy (\%) & 95\% CI (\%) & Diff.\ (pp), $p$ \\
    \hline
    Facial expression & 82.97 & 78.77--86.48 & $-13.46$, $1.4 \times 10^{-9}$ \\
    Speech emotion    & 85.71 & 81.75--88.94 & $-10.71$, $3.8 \times 10^{-7}$ \\
    Text-based        & 87.64 & 83.86--90.63 & $-8.79$, $1.4 \times 10^{-5}$ \\
    Multimodal        & \textbf{96.43} & \textbf{93.99--97.90} & -- \\
    \hline
  \end{tabular*}
\end{table}

The body movement component classified activity levels with an overall
accuracy of 94.87\% (Table~\ref{tab:body-results}).

\begin{table}[!t]
  \renewcommand{\arraystretch}{1.15}
  \caption{Accuracy of the Body Movement Component in the User Study}
  \label{tab:body-results}
  \centering
  \footnotesize
  \begin{tabular*}{\columnwidth}{@{\extracolsep{\fill}}lcc}
    \hline
    Activity level & Correct / total & Accuracy (\%) \\
    \hline
    Low      & 49 / 52   & 94.23 \\
    Medium   & 48 / 52   & 92.31 \\
    High     & 51 / 52   & 98.08 \\
    \hline
    Overall  & 148 / 156 & 94.87 \\
    \hline
  \end{tabular*}
\end{table}

\subsection{Comparison with Benchmark Performance}

The results reveal a clear gap between benchmark and realistic performance
for the single-modality components. The speech and text-based models, which
achieve 99.63\% and 97.96\% on their benchmark test sets
(Table~\ref{tab:model-summary}), reach 85.71\% and 87.64\% in the user study,
and the facial expression model falls from 85.21\% to 82.97\%. The multimodal
system, under the same conditions, reaches 96.43\%. The benchmark and user
study results are obtained in different ways, the former on labelled test
sets and the latter through participants' judgements of acted emotions, so
the comparison between them indicates a trend rather than an exact
difference. The comparison between the single-modality components and the
multimodal system within the user study, by contrast, is direct, as all were
evaluated by the same participants on the same emotional states.

% =====================================================================
\section{Discussion}
\label{sec:discussion}

\subsection{Principal Findings}

The findings highlight the limitations of single-modality emotion
recognition. High accuracy on a benchmark test set did not carry over to
realistic conditions, where each single channel was prone to
misclassifying particular emotional states. The multimodal system
compensated for these weaknesses, outperforming every single-modality
component and recognising all emotional states with an accuracy of at least
92.31\%. For a tool intended to support psychological evaluation, this
consistency across emotional states matters as much as overall accuracy: a
component that is accurate on average but fails on fear or on the neutral
state could mislead an evaluator precisely in the cases where support is
needed.

\subsection{Implications for Psychological Evaluation}

The system is intended as a supplementary tool that provides a quantitative,
independent reading of emotional state to psychologists, psychiatrists, and
other clinicians. By presenting the emotional profile alongside the overall
prediction, it allows the evaluator to see how confident the system is and
whether the channels agree, and to weigh this information against their own
observations. Its potential to mitigate human limitations such as fatigue,
bias, and inconsistent recall remains to be demonstrated in clinical use.

\subsection{Limitations}

The results should be interpreted in light of several limitations. The user
study was conducted remotely in a controlled setting with acted emotions,
designed to approximate a standardised interview, and the correctness of each
prediction was judged by the participants themselves. The study did not
involve clinicians or patients, and only 52 of the 159 individuals who began
the study completed it.

\subsection{Future Work}

The system should be evaluated by trained professionals with patients in
controlled clinical settings, to establish its value and safety in its
intended use. The body movement analysis could be extended from activity
level to movement features more directly associated with emotional state, and
eye tracking could be added to support the assessment of neurodevelopmental
conditions associated with social or attentional differences. The recognition
of weaker emotional states, such as fear for facial expressions, could be
improved through refined preprocessing or additional training data. Finally,
physiological signals, such as electroencephalography (EEG),
electrocardiography (ECG), heart rate variability (HRV), and galvanic skin
response (GSR), could provide further channels that are less subject to
voluntary control; these were not included in the present system owing to
cost.

% =====================================================================
\section{Conclusion}
\label{sec:conclusion}

This paper presented a multimodal emotion recognition system
intended to support psychological evaluation by providing a standardised,
data-driven reading of emotional state. The central finding is that
single-modality emotion recognition, even when it performs well on benchmark
test sets, falls short under realistic conditions, whereas combining
modalities substantially improves both overall accuracy and consistency
across emotional states, from 82.97--87.64\% for the single-modality
components to 96.43\% for the multimodal system. Because the system does not
depend on any single channel, multimodal fusion is expected to be
particularly valuable for individuals whose emotional expression is atypical,
such as those with neurodevelopmental conditions or limited facial
expressiveness, in whom a single channel may be absent or misleading.
Validating the system with these populations, and with clinicians and
patients more generally, is the most important direction for future work.

% =====================================================================
% Start the appendix on a new page so that its table is placed under the
% heading instead of floating into the references.
\clearpage
\appendix

Table~\ref{tab:fer-metrics} reports the per-class precision, recall, and F1
of the facial expression recognition models on their test sets.

\begin{table}[!h]
  \renewcommand{\arraystretch}{1.1}
  \caption{Facial Expression Recognition Model Performance Metrics. Macro
    averages are the unweighted mean of the per-class values; the overall
    figures reported in the text are weighted averages.}
  \label{tab:fer-metrics}
  \centering
  \footnotesize
  \begin{tabular*}{\columnwidth}{@{\extracolsep{\fill}}p{1.6cm}lccc}
    \hline
    Model & Class & Precision & Recall & F1 \\
    \hline
    \multirow{8}{1.6cm}{FER-2013}
      & Anger    & 54\%     & 54\%     & 54\%     \\
      & Disgust  & 64\%     & 43\%     & 52\%     \\
      & Fear     & 65\%     & 40\%     & 50\%     \\
      & Happiness & 85\%     & 92\%     & 88\%     \\
      & Neutral  & 57\%     & 81\%     & 67\%     \\
      & Sadness  & 49\%     & 25\%     & 33\%     \\
      & Surprise & 81\%     & 73\%     & 76\%     \\
      & Macro avg. & 65\%     & 58.29\%  & 60\%     \\
    \hline
    \multirow{8}{1.6cm}{RAF-DB}
      & Anger    & 88\%     & 78\%     & 83\%     \\
      & Disgust  & 84\%     & 60\%     & 70\%     \\
      & Fear     & 59\%     & 54\%     & 56\%     \\
      & Happiness & 91\%     & 96\%     & 94\%     \\
      & Neutral  & 80\%     & 79\%     & 79\%     \\
      & Sadness  & 89\%     & 72\%     & 79\%     \\
      & Surprise & 81\%     & 85\%     & 83\%     \\
      & Macro avg. & 81.71\%  & 74.86\%  & 77.71\%  \\
    \hline
    \multirow{8}{1.6cm}{CK+}
      & Anger    & 100\%    & 100\%    & 100\%    \\
      & Disgust  & 100\%    & 100\%    & 100\%    \\
      & Fear     & 99\%     & 99\%     & 99\%     \\
      & Happiness & 99\%     & 97\%     & 98\%     \\
      & Neutral  & 100\%    & 100\%    & 100\%    \\
      & Sadness  & 100\%    & 100\%    & 100\%    \\
      & Surprise & 100\%    & 100\%    & 100\%    \\
      & Macro avg. & 99.71\%  & 99.43\%  & 99.57\%  \\
    \hline
    \multirow{8}{1.6cm}{FER-2013 \& RAF-DB Combined}
      & Anger    & 59\%     & 60\%     & 60\%     \\
      & Disgust  & 69\%     & 48\%     & 57\%     \\
      & Fear     & 65\%     & 46\%     & 54\%     \\
      & Happiness & 85\%     & 95\%     & 90\%     \\
      & Neutral  & 62\%     & 79\%     & 70\%     \\
      & Sadness  & 81\%     & 53\%     & 64\%     \\
      & Surprise & 82\%     & 75\%     & 78\%     \\
      & Macro avg. & 71.86\%  & 65.14\%  & 67.57\%  \\
    \hline
  \end{tabular*}
\end{table}

% =====================================================================

\end{document}